\documentclass{article}
\usepackage{spconf,amsmath,graphicx}

\usepackage{enumitem}
\setlist{nosep, leftmargin=14pt}

\usepackage{mwe} 
\usepackage{graphicx}
\usepackage{todonotes}
\usepackage{bbding}
\usepackage{booktabs}
\usepackage{multirow,bigdelim}
\usepackage{color}
\usepackage{xr}

\usepackage{amsmath,amsfonts,amssymb}
\newcommand{\bftab}{\fontseries{b}\selectfont}
\usepackage{rotating}
\usepackage{mwe}
\usepackage{graphbox}
\usepackage{multirow}
\usepackage{amsfonts} 
\usepackage{amsmath,mathtools}
\usepackage{amssymb}
\usepackage{bm}
\usepackage{url}

\title{CATS: Complementary \underline{C}NN \underline{a}nd \underline{T}ransformer Encoders for \underline{S}egmentation}
\name{Hao Li, Dewei Hu, Han Liu, Jiacheng Wang, Ipek Oguz}
\address{anonymous}
\address{Department of Electrical Engineering and Computer Science, Vanderbilt University} 
\begin{document}
%

\maketitle
\begin{abstract}
Recently, deep learning methods have achieved state-of-the-art performance in many medical image segmentation tasks. Many of these are based on convolutional neural networks (CNNs). For such methods, the encoder is the key part for global and local information extraction from input images; the extracted features are then passed to the decoder for predicting the segmentations. In contrast, several recent works show a superior performance with the use of transformers, which can better model long-range spatial dependencies and capture  low-level details. However, transformer as sole encoder underperforms for some tasks where it cannot efficiently replace the convolution based encoder. In this paper, we propose a model with double encoders for 3D biomedical image segmentation. Our model is a U-shaped CNN augmented with an independent transformer encoder. We fuse the information from the convolutional encoder and the transformer, and pass it to the decoder to obtain the results. We evaluate our methods on three public datasets from three different challenges: BTCV, MoDA and Decathlon. Compared to the state-of-the-art models with and without transformers on each task, our proposed method obtains higher Dice scores across the board.
\end{abstract}
\section{Introduction}
In recent years, convolutional neural networks (CNNs) with U-shaped structures have dominated the medical image segmentation field \cite{cciccek20163d, schlemper2019attention, isensee2019automated}. The U-shaped networks consist of an encoder and a decoder, with skip connections in between. The encoder extracts information by consecutive convolution and down-sampling operations. The encoded information is sent to the decoder via skip connections to obtain the segmentation result. The U-Net \cite{cciccek20163d} and its many variants have shown great performance on many medical image segmentation tasks \cite{shapey2019artificial, hu2021life, li2021mri, zhang2021segmentation}.

However, convolutional encoders are somewhat limited for modeling long-range dependencies, due to the local receptive field of the convolution kernels. A potential solution is the transformer, which was originally proposed in the context of nature language processing, and has been used for image segmentation and classification \cite{dosovitskiy2020image, liu2021swin, xie2021cotr, valanarasu2021medical, shamshad2022transformers} due to its ability to better capture global information and modeling long-range context. For medical images, Chen et al.\ proposed the transUnet \cite{chen2021transunet}, which is the first segmentation framework with transformers. They added a transformer layer to the CNN-based encoder to better leverage global information. The UNEt TRansformers (UNETR) \cite{hatamizadeh2021unetr} is an architecture that completely replaces the CNN-based encoder with a transformer. The UNETR passes the multi-level scale information from this transformer to CNN-based decoder, and it is a state-of-the-art model for multi-organ segmentations on computed tomography (CT) images. Similar to U-Net, Cao et al.\ proposed Swin-Unet \cite{cao2021swin} which is a pure transformer-based U-shaped architecture for medical image segmentation. Since the transformer has limited ability to encode high-level information, these networks with transformers may not work well for some medical image segmentation tasks. To overcome this, Zhang et al.\ proposed a 2D architecture that combines a CNN-based encoder and a transformer-based segmentation network in parallel \cite{zhang2021transfuse}. 

In this paper, we propose CATS (Complementary \underline{C}NN \underline{a}nd \underline{T}ransformer Encoders for \underline{S}egmentation), a U-shaped architecture with double encoders. Inspired by UNETR \cite{hatamizadeh2021unetr} and TransFuse \cite{zhang2021transfuse}, we use a transformer as an independent encoder in addition to the CNN encoder. However, unlike the TransFuse, we use multi-scale features from the transformer rather than only the highest level features (i.e., the transformer output). The proposed CATS is a straightforward way to combine CNN and transformer without requiring a complex network architecture, such as the attention blocks and the BiFusion module in TransFuse. We use a 3D architecture and train from scratch instead of pre-training. Unlike the UNETR, we include both a transformer-based encoder and a CNN-based encoder. Multi-scale features extracted from the transformer are added with CNN features.  The fused information is delivered to the CNN-based decoder for segmentation. We compare our model to state-of-the-art models with and without transformers on three public datasets.

\begin{figure*}[t]
\centering
\includegraphics[width=\linewidth]{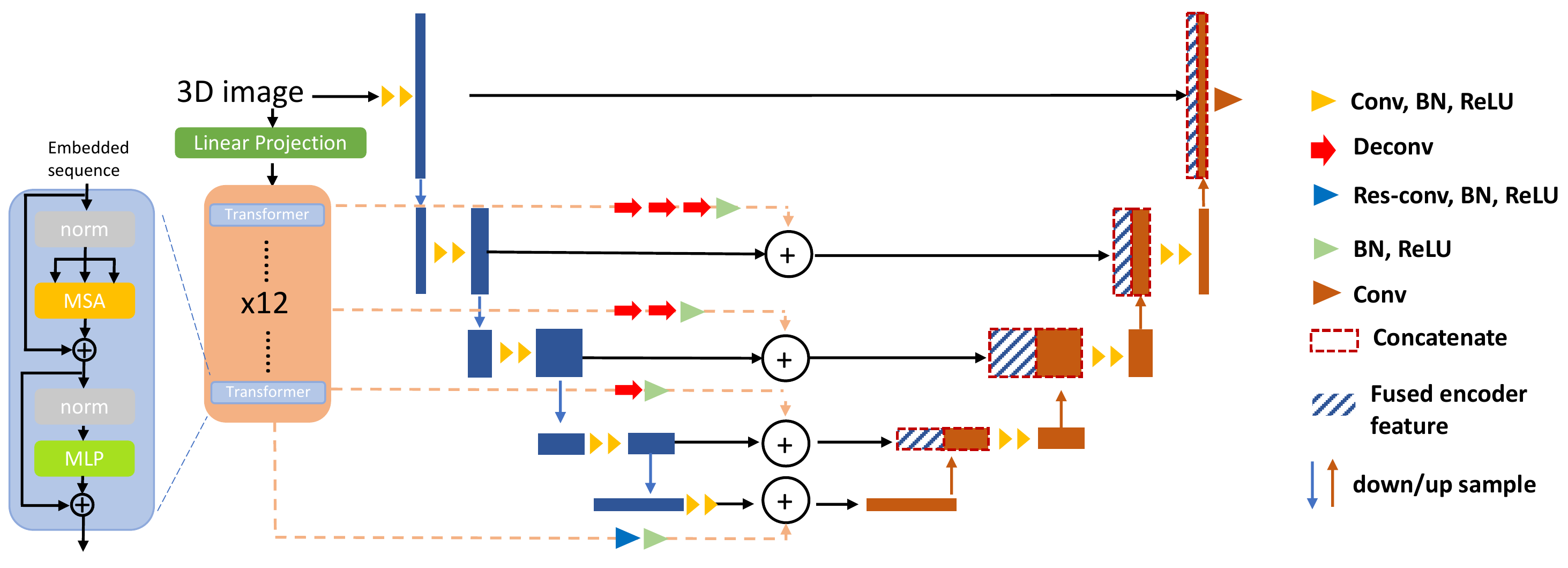}   
\caption{Proposed network architecture with two independent encoder paths: a CNN-based encoder and a transformer encoder.}
\label{network}
\end{figure*}

\section{Methods}
\subsection{Framework overview}
The proposed CATS framework is shown as Fig.~\ref{network}. Our model contains two encoder paths, a CNN path and a transformer path. For the CNN-based encoder, the information is gradually coded by the convolution and down-sampling operations. For the transformer path, to make sure the low-level details are well-preserved, we directly send the raw input to the transformer. Then, the information from the two paths are fused by addition operations at each level, and delivered to the CNN-based decoder to predict the final segmentation.

\subsection{Transformer}
We begin with an input 3D image $x\in\mathbb{R}^{C\times W \times H \times D}$, where $W$, $H$ and $D$ are the image dimensions and $C$ is the number of channels. We construct $x_p = \{x^i_p | i \in[1, N]\}$, which is a set consisting of $N$ non-overlapping patches $x^i_p \in \mathbb{R}^{(P^3\times C)}$, where $N = \frac{W \times H \times D}{P^3}$ and each $x^i_p$ is a patch of $P^3$ voxels with $C$ channels.
Next, we send the patches to the linear projection layer to obtain the embedded projection {$\bm{E}$} with dimension $M=P^3C$. To keep the position information, we add the position embedding $\bm{E_p}$ to form the transformer input:
\begin{equation}
z_0 = [x^1_p\bm{E};x^2_p\bm{E}; ... ;x^N_p\bm{E}] + \bm{E_p} 
\end{equation}
where $\bm{E}\in\mathbb{R}^{(P^3 \times C )\times M}$ and $\bm{E_p}, z_0 \in\mathbb{R}^{N\times M}$.

The encoder path of transformer (Fig.~\ref{network}) has $L$ layers of Multihead Self-Attention (MSA) and Multi-Layer Perceptron (MLP) blocks: 
\begin{equation}
z^{MSA}_l = MSA(norm(z_{l-1})) + z_{l-1}
\end{equation}
\begin{equation}
z_l=z^{MLP}_l = MLP(norm(z^{MSA}_l)) + z^{MSA}_l
\end{equation}
where $z^{MSA}_l$ and $z^{MLP}_l$ are the outputs from MSA and MLP blocks, $norm(\cdot)$ denotes the layer normalization and $l$ is the layer index.
The MLP blocks contain two linear layers followed by the GELU activation functions.

There are $n$ Self-Attention heads (SAs) in the MSA block to extract global information from the embedded sequence:
\begin{equation}
SA(z_i) = softmax(\frac{qk^T}{\sqrt{M_n}})v
\end{equation}
where $z_i\in\mathbb{R}^{N\times M}$, $q$, $k$ and $v$ are the query, key and value of $z_i$ respectively, $q = W_qz_i$, $k = W_kz_i$ and $v = W_vz_i$. $W$s are the three weight matrices and $\sqrt{M_n} = \frac{M}{n}$ is the scaling factor.
The output of the $softmax(\cdot)$ function is the similarity weight between $q$ and $k$.
Then the MSA is defined as:
\begin{equation}
MSA(z) = [SA_1(z);SA_2(z); ... ;SA_n(z)]W_{msa} 
\end{equation}
where $W_{msa}$ are the trainable weights.

Inspired by the UNETR \cite{hatamizadeh2021unetr}, we use the same strategy for visualizing the multi-scale features $z_3$, $z_6$, $z_9$, $z_{12}$ from transformer encoder path. The feature size of each $z_i$ is $N\times M = \frac{H}{P} \times \frac{W}{P}\times \frac{D}{P} \times M$. We upsample the $z_i$ into the same size as the corresponding outputs from CNN-based encoder by deconvolution (deconv) operations, batch normalization (BN) and RELU activation function. The details can be viewed in Fig.~\ref{network}. For the last level of the transformer features ($z_{12}$), we directly apply a residual convolution (res-conv) \cite{he2016deep}, BN and RELU for reshaping.

\subsection{Convolutional neural network architecture}
Our CNN (Fig.~\ref{network}) is adapted from the 3D U-Net \cite{cciccek20163d}. There are two parts of the CNN model, the encoder and the decoder. Four max-pooling and deconvolution operations are used for down-sampling and upsampling respectively. The feature maps from the top level are directly forwarded to the decoder, and the rest of the feature maps from the lower levels are fused with the encoded information from transformer path by addition. Then, the fused information is delivered to the decoder by skip connections that follows the way of 3D U-Net. 

\begin{table*}[t]
\caption{Mean Dice scores in BTCV dataset. Bold numbers denote the highest Dice scores. The results of TransUNet are directly copied from \cite{chen2021transunet}. The experiments of UNETR and proposed method use the public pipeline of UNETR. The organs from left to right are: spleen, right and left kidney, gallbladder, esophagus, liver, stomach, aorta, inferior vena cava, portal vein and splenic vein, pancreas, right and left adrenal gland, and overall average.}
\label{btcv}
\begin{center}
    \begin{tabular}{  l  c l l l l l l l l l l l l l l}
    \hline
    \hline
     Method & & Spl & RKid & LKid & Gall & Eso & Liv & Sto & Aor & IVC & Veins &Pan &RAG &LAG&Avg.\\
     \hline
     
     TransUNet \cite{chen2021transunet}& \vline& 85.1 & 77.0 & 81.9 & 63.1 & - & 94.1 & 75.6 & 87.2 & - & - 
     & 55.9 & - & - & 77.5 \\
     
    UNETR \cite{hatamizadeh2021unetr}& \vline& 93.4 & 85.5 & 87.6 & 61.9 & 74.7 & 95.7 & 76.8 & 85.2 & 77.2 &69.8 & 61.5 & 64.4 & 59.4 &76.9 \\
     
    CATS & \vline& \bftab95.8 & \bftab90.2 & \bftab93.4 & \bftab65.9 & \bftab77.1 & \bftab96.8 & \bftab83.0 & \bftab88.6 & \bftab83.1 & \bftab76.9 & \bftab73.8 & \bftab70.2 & \bftab62.6 & \bftab81.4\\
    \hline

  \end{tabular}
\end{center}
\end{table*}

\section{Experiments and Results}
\subsection{Datasets and Implementation Details}
\label{datasets}
We used three public datasets for our experiments, and followed the evaluation metrics of each challenge.

\textbf{Beyond the Cranial Vault (BTCV)}\footnote{\url{https://www.synapse.org/\#!Synapse:syn3193805/wiki/217753}} dataset contains 30/20 subjects with abdominal CT images for training/testing, with 13 different organs labeled by experts. 
To preprocess, we resampled the images and clipped HU values to range [-175, 250]. Three data augmentations were used: random flip, rotation and intensity shift. 
UNETR \cite{hatamizadeh2021unetr} is the winner for this challenge, and we compared our results to the publicly available UNETR implementation 
\footnote{\url{https://github.com/Project-MONAI/tutorials/blob/master/3d_segmentation/unetr_btcv_segmentation_3d.ipynb}}. We also compare to the TransUNet model \cite{chen2021transunet}. We note that both of these methods have been shown to be superior to CNN-only models such as 3D U-Net for this challenge 
\cite{chen2021transunet, hatamizadeh2021unetr}. 

\textbf{Cross-Modality Domain Adaptation for Medical Image Segmentation (MoDA)}\footnote{\url{https://crossmoda.grand-challenge.org/}} has 105 contrast-enhanced T1-weighted MRIs with manual labels for vestibular schwannomas (VS). We split the dataset into 55/20/30 for training/validation/ testing.  The preprocessing pipeline consists of rigid registration to MNI space and cropping.
Adam optimizer and Dice loss are used. We again compare our results to  TransUNet\cite{chen2021transunet} and UNETR \cite{hatamizadeh2021unetr}, as well as to a 2.5D CNN \cite{shapey2019artificial} which was provided as the baseline for the challenge. We also report the 95\% Hausdorff distance in this dataset in addition to the metrics used in the challenge.

\textbf{Task 5 of Decathlon (Decathlon-5)}\footnote{\url{http://medicaldecathlon.com/}}  consists of 32 MRIs with manual prostate labels. 2 MRIs in validation were excluded due to the wrong labels being provided in the public dataset.
We compare to the TransFuse \cite{zhang2021transfuse} model and the nn-Unet \cite{isensee2019automated}, which was the top-performing approach of the challenge at the time of the initial competition for this task. For a direct comparison with these two models, we use this dataset in a 5-fold cross-validation framework, and follow the setting in \cite{isensee2019automated}.

\textbf{Implementation details.} The training batch size was 2 for all three experiments, and constant learning rate was 0.0001. All intensities were normalized to range [0, 1]. We used Pytorch, MONAI and an Nvidia Titan RTX GPU.

\subsection{BTCV Results}

\begin{figure}[t]
\centering
\includegraphics[width=\linewidth]{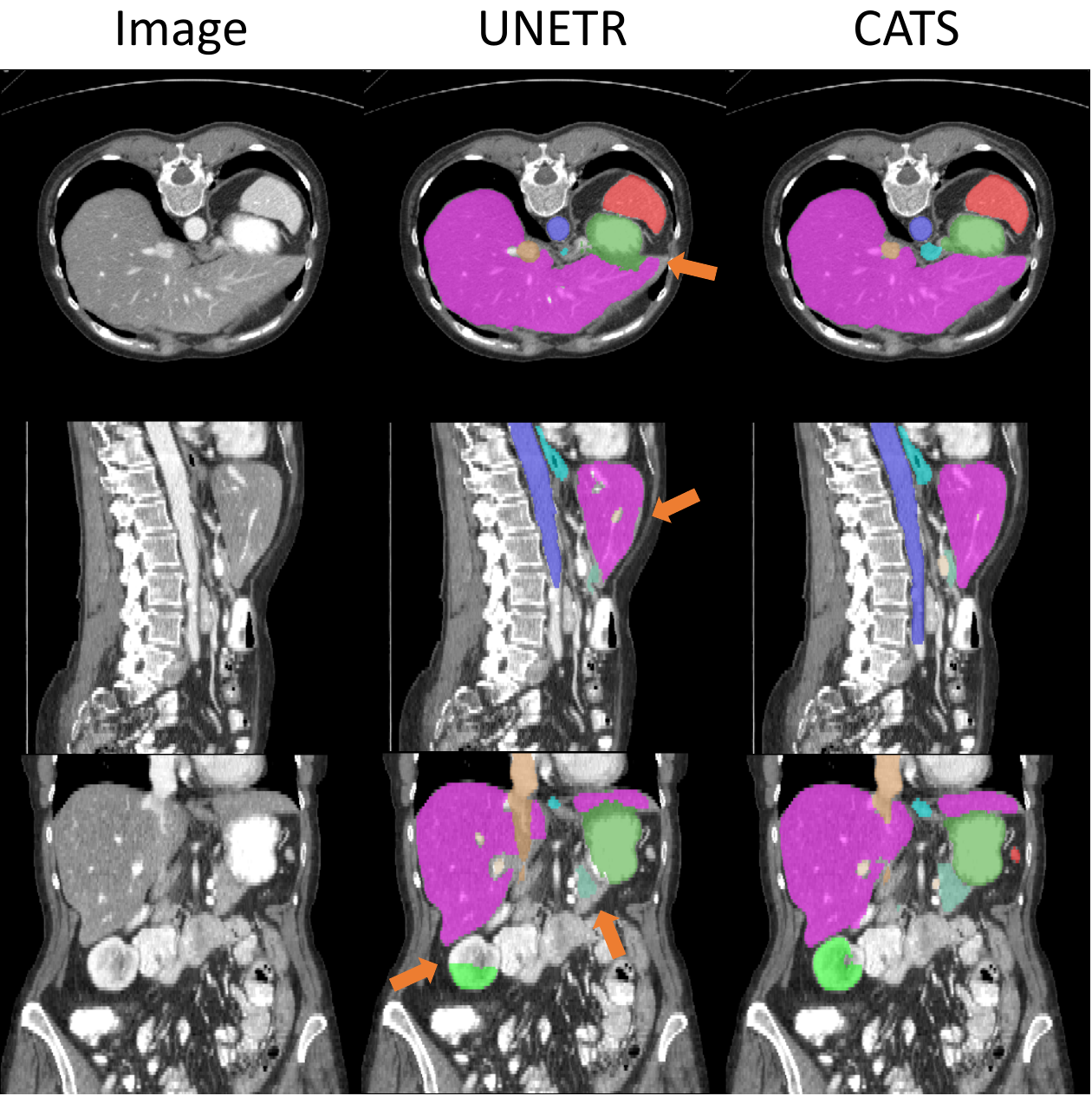}
\caption{Qualitative results in BTCV test set. Some major differences are highlighted by orange arrows.}
\label{BTCV}
\end{figure}

\begin{figure*}[h]
\centering
\includegraphics[width=\linewidth]{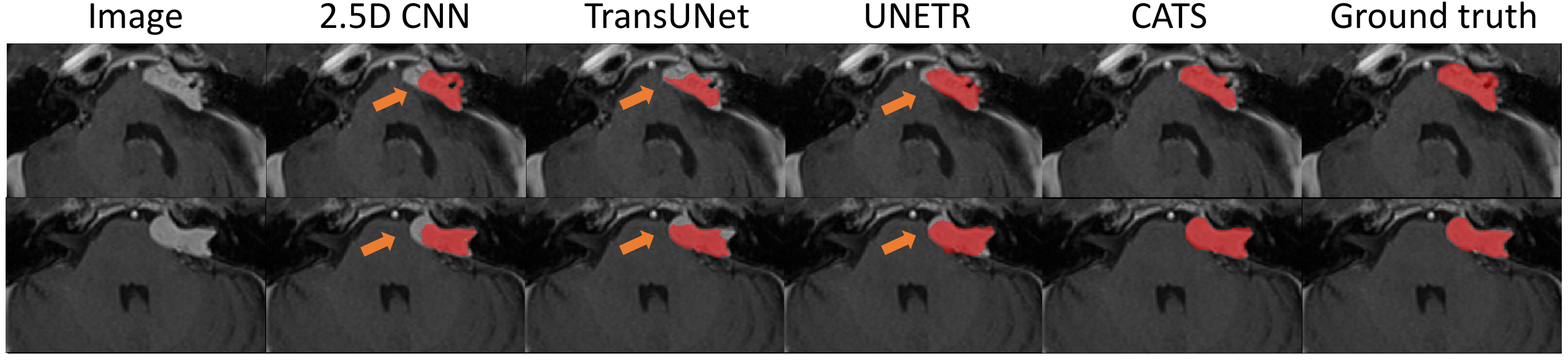}   
\caption{Quantitative results in MoDA. Local segmentation errors are highlighted with arrows.}
\label{moda123}
\end{figure*}

The Dice score is used to evaluate the BTCV experiment; the results can be viewed in Tab.~\ref{btcv}. Our proposed method outperformed the state-of-the-art transformer-based models for each organ in this dataset. The most dramatic improvements (Dice improvement $ > 5\%$) between UNETR and our proposed method are in left kidney, stomach, inferior vena cava, portal vein and splenic vein, pancreas and right adrenal gland. It is noteworthy that our method improves the segmentation accuracy not only on larger organs such as stomach and kidney, but also on small organs such as the right adrenal gland.

Fig.~\ref{BTCV} shows qualitative results. Compared to the UNETR, our proposed model produces smoother segmentation for the stomach and liver (axial view, arrow). We can also see (coronal view, arrows) that UNETR undersegments the right kidney and liver, unlike our proposed model.

\begin{table}[b]
\caption{Quantitative results in MoDA dataset, presented as $mean (std.dev.)$. Bold numbers indicate the best performance.}
\label{moda}
\begin{center}
    \begin{tabular}{  l  c c c c}
    \hline
    \hline
     Method & & Dice & ASD & HD95\\
     \hline
     
     2.5D CNN \cite{shapey2019artificial} & \vline& 0.856 (1.000) & 0.69 (1.20) & 3.5 (5.2) \\
     
     TransUNet \cite{chen2021transunet} & \vline& 0.792 (0.234) & 7.86 (27.6) & 12 (31) \\
     
     UNETR \cite{hatamizadeh2021unetr}& \vline& 0.772 (0.139) & 7.95 (14.2) & 26 (43) \\
     
     CATS & \vline& \bftab0.873 (0.088) & \bftab0.48 (0.63) & \bftab2.6 (3.6) \\
    \hline

  \end{tabular}
\end{center}
\end{table}


\subsection{MoDA results}

The quantitative results of MoDA dataset are shown in Tab.~\ref{moda}. We report the Dice score, average surface distance (ASD) and 95-percent Hausdorff distance (HD95) as metrics. It is easy to observe that the CNN-only network has better performance than the  transformer-only encoders for this task. However, our proposed CATS model outperformed the 2.5D CNN \cite{shapey2019artificial}, which was specifically designed for segmenting VS from MRIs with large difference between in-plane resolution and slice thickness, as is the case for this dataset.

Fig.~\ref{moda123} shows the qualitative results of VS segmentation. While all methods appear to undersegment the VS, our proposed model most closely resembles the ground truth segmentations.

\begin{table}[t]
\caption{Mean Dice scores in Decathlon-5 dataset. PZ and TZ denote the peripheral zone and the transition zone of the prostate, respectively.}
\label{prostate}
\begin{center}
    \begin{tabular}{  l  c c c c}
    \hline
    \hline
     Method & & PZ & TZ & Avg.\\
     \hline
     2D nnUnet \cite{isensee2019automated} & \vline& 0.6285 & 0.8380 & 0.7333 \\
     
     3D nnUnet full \cite{isensee2019automated}& \vline& 0.6663 & 0.8410 & 0.7537 \\
     
     TransFuse-S \cite{zhang2021transfuse} & \vline& 0.6738 & 0.8539 & 0.7639 \\
     
     CATS & \vline& \bftab0.7136 & \bftab0.8618 & \bftab0.7877 \\

    \hline

  \end{tabular}
\end{center}
\end{table}

\subsection{Decathlon-5 results}
We compare the nnUnet and TransFuse for the prostate segmentation in Tab.~\ref{prostate}. Our proposed method has the highest Dice scores on all labels. Moreover, we improved the peripheral zone (PZ) nearly $4\%$ compared to the performance of the TransFuse model.

\section{Discussion and Conclusions}
In this paper, we propose a convolutional neural network with a transformer as an independent encoder. The transformer can complement the CNN by modeling long-range dependencies and capturing low-level details. We evaluate our proposed method on three public datasets: (1) BTCV, (2) MoDA and (3) Decathlon. Compared to the state-of-the-art models which also attempt to incorporate transformers into the segmentation networks in various ways, our proposed model has superior performance on each task. We believe this is due to our efficient integration of the transformer and CNN encoders, as well as our use of a 3D architecture compared to 2D models. In future work, we will apply our method to larger public datasets. Additionally, others transformer layers may be helpful to further improve the performance. 

\section{Compliance with Ethical Standards}
This research study was conducted retrospectively using human subject data made available in open access by BTCV, MoDA and Decathlon (see detailed information in Sec.~\ref{datasets}). Ethical approval was not required as confirmed by the license attached with the open access data.

\section{Acknowledgments}
This work was supported by NIH grant R01-NS094456.

\bibliographystyle{IEEEbib}
\bibliography{refs}

\end{document}